\documentclass[pra,twocolumn]{revtex4-2}
\usepackage{float,graphicx,multirow, amsmath,color,dsfont}
\usepackage{amsmath,bm}
\usepackage{amssymb,mathrsfs,dsfont}
\usepackage{amssymb,mathbbol}
\usepackage{amsfonts}
\usepackage{mathrsfs}
\usepackage{float}
\usepackage{xcolor,hyperref}
\definecolor{darkblue}{rgb}{0,0.0.1,0.3}
\definecolor{darkred}{rgb}{0.6,0.1,0}
\hypersetup{colorlinks,breaklinks,
	linkcolor=darkred,urlcolor=darkblue,
	anchorcolor=darkred,citecolor=
	darkred,pdfauthor=SaAr, pdftitle=SaAr.HBT}
\usepackage{tikz}
\usepackage{mathtools, physics}
\usepackage{float}
\usepackage{multirow,bigdelim}
\usepackage{subfig}

\usepackage[normalem]{ulem}

\begin{document}

\title{Violation of Bell-type Inequality in the Hanbury Brown-Twiss
(HBT) Toy-Model: Quantum Correlations beyond Local-Realism}

\author{Sarbani Chatterjee}
\email{mp18015@iisermohali.ac.in}

\author{Arvind}
\email{arvind@iisermohali.ac.in}
\affiliation{Department of Physical Sciences, Indian Institute
of Science Education and Research Mohali, Sector 81 SAS Nagar, 
Punjab 140306 India.}

\date{\today}

\begin{abstract}
The Hanbury Brown–Twiss (HBT) effect marked a significant 
milestone in quantum optics by illustrating that intensity 
correlations can reveal inherently quantum features of light. 
Motivated by the structure of the original $8$-mode HBT 
interferometer, we develop a simplified $4$-mode toy-model that 
captures its essential physical ingredients while significantly 
reducing analytical complexity. Our toy-model provides an 
intuitive and well-structured framework for examining the 
desired correlation functions. We investigate whether our 
toy-model can exhibit genuinely non-classical behaviour by 
formulating and applying a Bell-type inequality tailored to its 
specific measurement configuration. Our analysis shows that the 
correlations generated by the toy-model violate the inequality, 
indicating the presence of non-local features that are 
incompatible with any classical hidden-variable description. 
The results highlight the power of simplified models in 
uncovering fundamental aspects of non-classicality in optical 
systems.
\end{abstract}

\maketitle

\section{Introduction}
\label{sec:1_intro}
The initial years of the field of `quantum optics' primarily
involved studying laser physics, where the quantum aspects
were linked only to the amplifying medium, and light itself
was described classically as an electromagnetic
wave~\cite{lamb_1964}. Until the 1960s, spontaneous emission
was the only phenomenon that required a quantum description,
while other processes could still be explained using
semi-classical models~\cite{Mandel_Wolf_1995,
Grynberg_Aspect_Fabre_2010}. It was only in the 1960s that
the idea of quantizing freely propagating light emerged,
leading to the conception of photons and a broader
understanding of quantum optics.

The Hanbury Brown–Twiss (HBT) effect, discovered in 
1956~\cite{hbt_1956,hbt_1956_2}, was a revolutionary 
experiment that was the first to successfully measure 
correlations between the intensity fluctuations of light 
arriving at two separate detectors, leading to the introduction 
of intensity interferometry. Originally developed to measure 
the angular size of stars~\cite{hbt_astro_1974, foellmi_2009}, 
the HBT effect was examined in both spatial and temporal 
domains. Examining intensity correlations revealed that light 
exhibits quantum statistical behaviour, such as ``photon anti-
bunching''~\cite{Loudon_1976, mandel_1977, rom_2007}, 
which could not be explained by classical physics. Although 
initially there was dubiety with regard to the quantum 
mechanical meaning of HBT correlations, this effect formally 
laid foundations to the field of quantum optics. Along with the 
Hong-Ou-Mandel (HOM)~\cite{hom_1987} and other similar 
effects~\cite{mandel_1977, grangier_1986, aspect_2019}, it led 
to the second quantum revolution~\cite{dowling_2003, 
Bell_Aspect_2004} and subsequently developments in modern 
quantum technologies, including quantum imaging, cryptography 
and information science among others~\cite{campagnano_2012, 
Munoz_2016, arvind_2017, capolupo_2020}.

In this work, we propose a simplified toy-model for the HBT 
experimental setup in the spatial regime. Our model reduces the 
original $8$-mode setup to a $4$-mode quantum analog keeping 
all its physical properties intact while significantly reducing 
analytical complexity. We also demonstrate that our toy-model 
is a valid candidate for explaining the intricacies of the HBT 
setup and that it is consistent with corresponding local hidden 
variable theories.

Building on the famous Einstein–Podolsky–Rosen (EPR) paradox
of 1935~\cite{epr_1935}, John S.~Bell formulated his inequality 
in 1964~\cite{bell_1964}, according to which any theory based 
on local hidden variables must satisfy certain statistical 
inequalities. Subsequently, several experimental 
tests~\cite{chsh_1969, clauser_1978, aspect_1982, 
aspect_grangier_1982, zeilinger_1999} were conducted on these 
inequalities which found clear violations, confirming the 
predictions of quantum theory and establishing the existence of 
`quantum non-locality' as a real physical phenomenon. 

In quantum optics, states can be characterized by
a positive Sudarshan-Glauber $P$ function and thus allow an
interpretation in terms of classical probability distribution 
over coherent states, which are known to satisfy Bell’s 
inequalities, whereas non-classical states can 
potentially lead to their violations. Such inequalities 
delineate the limits of local realistic extensions of quantum 
mechanics and can be violated by certain entangled quantum 
states~\cite{cohen_1997, arvind_1999, huntington_2001, 
chandan_2021}.

In this work, we study specific quantum states of radiation, 
particularly squeezed vacuum states, in the HBT scenario that 
can produce correlations which exceed the classical theoretical 
limits. It should be emphasized that our results, however, do 
not undermine the fundamental classical significance of HBT 
correlations.  Our work instead focuses on specific quantum 
states of light that are capable of producing correlations 
stronger than any classical theory would allow.

This paper is structured as follows: 
Section~\ref{sec:2_hbt_effect} discusses the original HBT 
effect, with background on the $4$-point correlation function. 
We introduce our proposed HBT toy-model in 
Sec.~\ref{sec:2_2_hbt_toy_model}, with focus on its similarity to 
the original HBT experimental setup, and details on the 
construction of our toy-model in 
Sec.~\ref{sec:3_hbt_toy_model_detail}. Mathematical 
formulation for the preparation of our resource state is 
discussed in Sec.~\ref{sec:3_3_hbt_toy_model_state}. In 
Sec.~\ref{sec:4_bell_inequality}, we derive the Bell-type 
inequality with dichotomous observables, that we use in our 
work, and report the obtained non-local features in 
Sec.~\ref{sec:5_results}. We summarize our results in 
Sec.~\ref{sec:6_conclusion}. Appendix~\ref{sec:7_appendixA} 
provides additional details on continuous variable (CV) systems 
and symplectic transformations used in our work.

\section{The Hanbury Brown-Twiss (HBT) effect}
\label{sec:2_hbt_effect}

The HBT effect was originally designed as an intensity 
interferometry technique for measuring the angular diameter of 
distant stellar sources. Prior to the development of this 
effect, interference of optical fields were measured using 
Michelson interferometer which required precise phase stability 
in order to minimize signal distortions, communication errors 
and measurement inaccuracies caused by phase shifts. Unlike 
this conventional interferometer, the HBT interferometer 
focuses on correlations between intensity fluctuations measured 
at two spatially separated detectors, making it inherently 
insensitive to phase fluctuations.

\subsection{HBT in an astronomical context}
\label{sec:hbt_astro}

Consider two localized, quasi-monochromatic radiation sources 
$S_1$ and $S_2$, located at a distance $s$ apart. Light from 
each source reaches two photo-detectors $D_1$ and $D_2$, which 
are placed distance $d$ away from each other at points 
$\bm{r}_1$ and $\bm{r}_2$ respectively. Light waves 
propagating from the sources travel through the distance $L$ to 
reach the detectors, with $s,\, d << L$. This assumption allows 
us to practically treat these propagation vectors to be 
parallel to each other, i.e. we can now work in the paraxial 
regime.

\begin{figure}[h]
    \centering
    \includegraphics[width=1\linewidth]{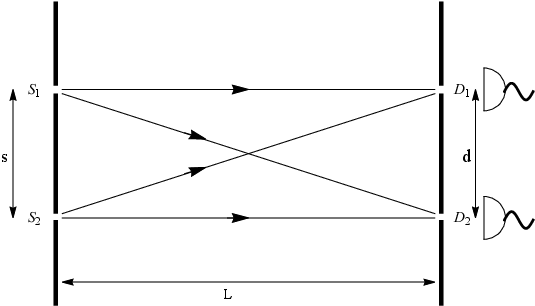}
    \caption{Typical Hanbury Brown-Twiss
    (HBT) experimental setup with two quasi-monochromatic 
    sources ($S_1,\, S_2$) and two photo-detectors 
    ($D_1,\, D_2$) with the assumption $s,\, d << L$.}
    \label{fig:hbt_classical}
\end{figure}

From the semi-classical theory of quantum optics, the 
photocurrent $i$, measured by a photo-detector at position 
$\bm{r}_j$, is proportional to the intensity of light $I$ 
incident on it, which in terms of the squared modulus of the 
classical complex electric field $\bm E$ is

\begin{equation}
    \label{eqn:hbt_photocurrent}
    i(\bm{r}_j) \; \propto \; I(\bm{r}_j) \; = \; 
    |\bm E(\bm{r}_j)|^2 \; = \; \bm E^*(\bm{r}_j) \; \bm E(\bm{r}_j).
\end{equation}

Using the above formula, the intensity-intensity correlation 
between the measurement outputs of $D_1$ and $D_2$ of the HBT 
setup in Fig.~\ref{fig:hbt_classical} is then~\cite{glauber_1963, arvind_2017}

\begin{equation}
    \begin{split}
        \label{eqn:hbt_correlation_function}
        G^{(2)}(r) &=  \;  
        \langle \; I_1(\bm{r}_1) \; I_2(\bm{r}_2) \;\rangle \\
        &= \; \langle \; \bm E_1^*(\bm{r}_1) \; 
        \bm E_1(\bm{r}_1) \; \bm E_2^*(\bm{r}_2) \;
        \bm E_2(\bm{r}_2) \; \rangle.
    \end{split}
\end{equation}

$G^{(2)}(r)$ is also called the HBT
correlation function or the $4$-point correlation function of 
the classical complex electric field, with $I_j(\bm{r}_j)$ 
and $\bm E_j(\bm{r}_j)$ as intensities and electric fields 
at the two detectors, respectively, for $j = 1,\, 2$. Upon 
normalization, Eq.~\eqref{eqn:hbt_correlation_function} becomes

\begin{equation}
        \label{eqn:hbt_correlation_norm}
        g^{(2)}(r) = \frac{\langle 
        I_1(\bm{r}_1) I_2(\bm{r}_2) \rangle}{\langle 
        I_1(\bm{r}_1)\rangle \langle I_2(\bm{r}_2) 
        \rangle}.
\end{equation}

The first-order spatial coherence term, $g^{(1)}(r)$, defined as~\cite{glauber_1963}

\begin{equation}
        \label{eqn:hbt_g1_correlation_norm}
        g^{(1)}(r) = \frac{\langle 
        \bm E_1^*(\bm{r}_1) \; \bm E_2(\bm{r}_2)
        \rangle}{\sqrt{\langle \bm E_1^*(\bm{r}_1) \bm E_1(\bm{r}_1)
        	 \rangle \; \langle \bm E_2^*(\bm{r}_2) \bm E_2(\bm{r}_2) \rangle}},
\end{equation}
measures the amplitude-amplitude correlations of the field to 
quantify spatial coherence. For a thermal/chaotic source 
characterized by random, independent and uncorrelated motion 
of its atoms, the intensity correlation $(g^{(2)})$ is related 
to the normalized first-order spatial coherence term 
$(g^{(1)})$ through the 
Siegert relation~\cite{siegert_1943}, $g^{(2)}(r) = 
1+\left|g^{(1)}(r)\right|^2$. Now, the 
van Cittert–Zernike theorem~\cite{zernike_1938} states that $g^{(1)}$ is equivalent 
to the normalized Fourier transform of the source's angular 
brightness distribution. Thus, by measuring $g^{(2)}$ with 
different detector separations, the spatial extent of the 
emitting source can be directly deduced~\cite{foellmi_2009, ferreira_2020}.

Consider, for example, an uniformly illuminated stellar disk of 
angular diameter $\theta$, then the spatial coherence is given by 
\begin{equation}
        \label{eqn:hbt_star}
        \left|g^{(1)}(d)\right|=\frac{2J_1\left(\pi d \theta/
        \lambda\right)}{\pi d \theta/\lambda},
\end{equation}
where $d$ is the distance (baseline) between the two detectors, 
$\lambda$ is the wavelength of detected light, and $J_1(\cdot)$ 
is the first-order Bessel function. Thus, by monitoring the 
decay of intensity correlations across an increasing baseline, 
the angular diameter of a star can be directly determined.

Beyond its astrophysical origin, the HBT effect further
established the foundations for modern quantum optics and 
temporal photon statistics. Unlike the HBT setup 
explained above, in a generalized quantum optics 
configuration, measurements are 
taken by the two detectors placed at positions $\bm r_1$ and 
$\bm r_2$ at separate times $t_1$ and $t_2$ respectively, where 
$t_2 = t_1 + \tau$, i.e., the second measurement is taken after 
a time delay. This change in the setup demonstrates that second-
order coherence provides information about the statistical 
nature of light that is otherwise inaccessible through the first-
order interference alone, as explained below. It turns out 
that within classical description there are constraints on the 
values that $g^{(2)}$ function can take. These constraints are 
no longer valid when light is described quantum mechanically and 
violation of such constraints happens in situations which are
intrinsically quantum in nature. The HBT experiment thus proves 
to be a benchmark for identifying intrinsically non-classical
situations in the context of light.

\subsection{The $8$-mode HBT configuration}
\label{sec:hbt_8mode}

The most general quantum mechanical form of the HBT 
correlation function in terms of both space and time can be 
written as~\cite{glauber_1963, Mandel_Wolf_1995}:
    \begin{align}
        \label{eqn:hbt_time_space}
        g^{(2)}(\tau; r) &= \frac{\langle : \hat{I}_1(t_1,
        \bm r_1) \hat{I}_2(t_2, \bm r_2) : \rangle}{\langle 
        \hat{I}_1(t_1, \bm r_1) 
        \rangle \langle \hat{I}_2(t_2, \bm r_2)   \rangle}, \\
        &= \frac{\langle : \bm E_1^*(t_1, \bm r_1) \bm E_2^*(t_2,
        \bm r_2) \bm E_2(t_2, \bm r_2) \bm E_1(t_1, \bm r_1) : 
        \rangle}{\langle \bm E_1^*(t_1,\bm r_1) \bm E_1(t_1, \bm 
        r_1) \rangle \langle \bm E_2^*(t_2,\bm r_2) \bm E_2(t_2, 
        \bm r_2)\rangle},\notag
    \end{align}
\noindent where, $r = | \bm r_1 - \bm r_2 |$, $\tau = t_1 - t_2$ 
and $ : \cdot : $ represents normal ordering of the operators.

For the HBT setup in Fig~\ref{fig:hbt_classical},
for each of 
the four propagation directions we can imagine a 
$2$-dimensional polarization space with two mutually 
orthogonal polarizations -- horizontal (H) and vertical (V)
providing a basis set, 
which essentially makes the system an $8$-mode interferometer. 
Each of these modes can be associated to creation and 
annihilation operators, $\hat{a}^{\dagger}_{ij,H(V)}$ and 
$\hat{a}_{ij,H(V)}$, where $i, \; j = 1,2$ for the two sources 
$S_i$ and detectors $D_j$. The HBT correlation function in 
Eq.~\eqref{eqn:hbt_time_space} is written in terms of mode 
creation and annihilation operators as follows:

\begin{equation}
\begin{split}
    \label{eqn:hbt_8mode}
     g^{(2)}(\tau; r) = \hspace{4cm} \\
     \frac{ \splitfrac{\langle : (\hat{a}_{11,H}^{\dagger} 
     \hat{a}_{11,H} + \hat{a}_{11,V}^{\dagger} \hat{a}_{11,V} + 
     \hat{a}_{21,H}^{\dagger} \hat{a}_{21,H} + \hat{a}_{21,V}
     ^{\dagger} \hat{a}_{21,V})}{(\hat{a}_{12,H}^{\dagger} 
     \hat{a}_{12,H} + \hat{a}_{12,V}^{\dagger} \hat{a}_{12,V} + 
     \hat{a}_{22,H}^{\dagger} \hat{a}_{22,H} + \hat{a}_{22,V}
     ^{\dagger} \hat{a}_{22,V}) : \rangle}}{\splitfrac{\langle 
     (\hat{a}_{11,H}^{\dagger} \hat{a}_{11,H} + \hat{a}_{11,V}
     ^{\dagger} \hat{a}_{11,V} + \hat{a}_{21,H}^{\dagger} \hat{a}
     _{21,H} + \hat{a}_{21,V}^{\dagger} \hat{a}_{21,V}) \rangle}
     {\langle (\hat{a}_{12,H}^{\dagger} \hat{a}_{12,H} + \hat{a}
     _{12,V}^{\dagger} \hat{a}_{12,V} + \hat{a}_{22,H}^{\dagger} 
     \hat{a}_{22,H} + \hat{a}_{22,V}^{\dagger} \hat{a}
     _{22,V})\rangle}}.
\end{split}
\end{equation}

It should be noted that measurements at the two detectors placed 
at $\bm r_1$ and $\bm r_2$, are recorded at times $t_1$ and 
$t_2$ respectively, i.e., $\tau$ is the time difference between 
the two measurements and $r$ is the path difference acquired 
by the light beams while reaching the detectors.

By keeping the two sources and two detectors in place, but 
restricting the number of wave propagation directions to two 
instead of four, we end up with our toy-model with 
$4$-polarization modes that we propose in the next 
subsection. The additional 
functionalities of the original $8$-mode HBT configuration are 
implemented by integrating various optical components between 
the sources and detectors.

\subsection{The $4$-mode HBT toy-model}
\label{sec:2_2_hbt_toy_model}

In the previous section, we described a
typical HBT setup. We now propose a toy-model which while adhering to the essential aspects
considerably simplifies the analysis, particularly 
for a quantum mechanical treatment.

\begin{figure*}
    \centering
    \includegraphics[width=\textwidth]{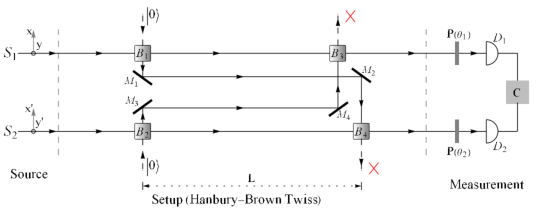}
    \caption{Schematic of our proposed HBT $4$-mode toy-model,
    inspired from the original $8$-mode setup, that allows 
    simplified theoretical analysis of the system while keeping 
    the essential physical ingredients of the HBT effect intact.}
    \label{fig:hbt_toy_model}
\end{figure*}

For our HBT toy-model, shown in Fig.~\ref{fig:hbt_toy_model}, 
we consider light emerging from sources $S_1$ and $S_2$ to 
propagate along $z$ and $z'$ directions respectively. The 
corresponding polarization modes are thus along the $x, y$ and 
$x', y'$ axes. We strategically position optical elements 
along the trajectories of these two beams in order to mimic 
their paths in Fig.~\ref{fig:hbt_classical}. The optical 
elements used in our construction include beam splitters 
($B_k$), mirrors ($M_k$) and polarizers ($P(\theta_m)$), with 
$k = 1, \dots, 4$ and $m = 1, 2$. The two beams emerging out 
of the HBT toy-model are then measured by `on-off' detectors 
$D_1$ and $D_2$ after passing through linear polarizers 
$P(\theta_1)$ and $P(\theta_2)$ respectively.

Note that in our toy-model description, light from each source 
possesses two modes, making our 
toy-model an effective $4$-mode interferometer. Although our 
model integrates four extra modes at the ports of 
beam splitters $B_1$ and $B_2$, they are effectively excluded 
after mixing of the beams at $B_3$ and $B_4$. Thus, our 
simplified $4$-mode toy-model, inspired from the 
structure of the original $8$-mode HBT setup, significantly 
reduces mathematical complexities.

Since we are restricting our study to the Schr\"odinger 
picture, the effects of the polarizers are incorporated into the quantum state, while the mode operators remain unaffected. Therefore, the bosonic commutation relation between the modes still holds, i.e., $[\hat{b}_{j,k}, \hat{b}_{j',k'}^{\dagger}] = \delta_{jj'} \delta_{kk'}$, where $j = 1, 2$ denote the two detectors and $k = H, V$ corresponds to the horizontal and vertical polarization directions respectively. Let us now consider a commutation relation defined in terms of the number operator $ [ \hat{n}_{j,k} , \hat{n}_{j',k'} ]$, where $\hat{n}_{j,k} = \hat{b}_{j,k}^{\dagger} \hat{b}_{j,k}$.
Since the average photon number is directly proportional to intensity of light incident on the detectors which is a key parameter for the HBT correlation function, the commutation relation has been formulated in terms of the photon number operators. It can be easily seen that under the conditions $j \ne j'$, i.e., for two different propagation directions, the number operators commute with each other. This indicates that a photon number measurement at $D_1$ does not disturb the measurement outcome at $D_2$, and are therefore simultaneously observable. Without loss of generality, the photo-detectors in the HBT setup can be replaced by `on-off' detectors. The remaining part of this work shall explicitly focus on measurements using `on-off' detectors.

For the $4$-mode toy-model, the HBT correlation function in Eq.~\eqref{eqn:hbt_time_space} takes the form:
\begin{equation}
    \begin{split}
        g^{(2)}(\tau; r) =  \hspace{3cm} \\
 \frac{\langle : (\hat{b}_{1,H}^{\dagger} \hat{b}_{1,H} + 
     \hat{b}_{1,V}^{\dagger} \hat{b}_{1,V}) \; (\hat{b}_{2,H}
     ^{\dagger} \hat{b}_{2,H} + \hat{b}_{2,V}^{\dagger} \hat{b}
     _{2,V}) : \rangle}{\langle (\hat{b}_{1,H}^{\dagger} \hat{b}
     _{1,H} + \hat{b}_{1,V}^{\dagger} \hat{b}_{1,V}) \rangle \; 
     \langle (\hat{b}_{2,H}^{\dagger} \hat{b}_{2,H} + \hat{b}
     _{2,V}^{\dagger} \hat{b}_{2,V}) \rangle}.
\label{eqn:hbt_4mode}
    \end{split}
\end{equation}

The minimum number of modes that has to be 
considered to study the HBT effect assuming two distinct sources 
is thus four. Further reducing the number of modes limits the 
setup to a single source, which still is a valid HBT
configuration for auto-correlations studies.

\subsection{The single-mode HBT configuration and photon
statistics}
\label{sec:hbt_1mode}

\begin{figure}[h]
    \centering
    \includegraphics[width=1\linewidth]{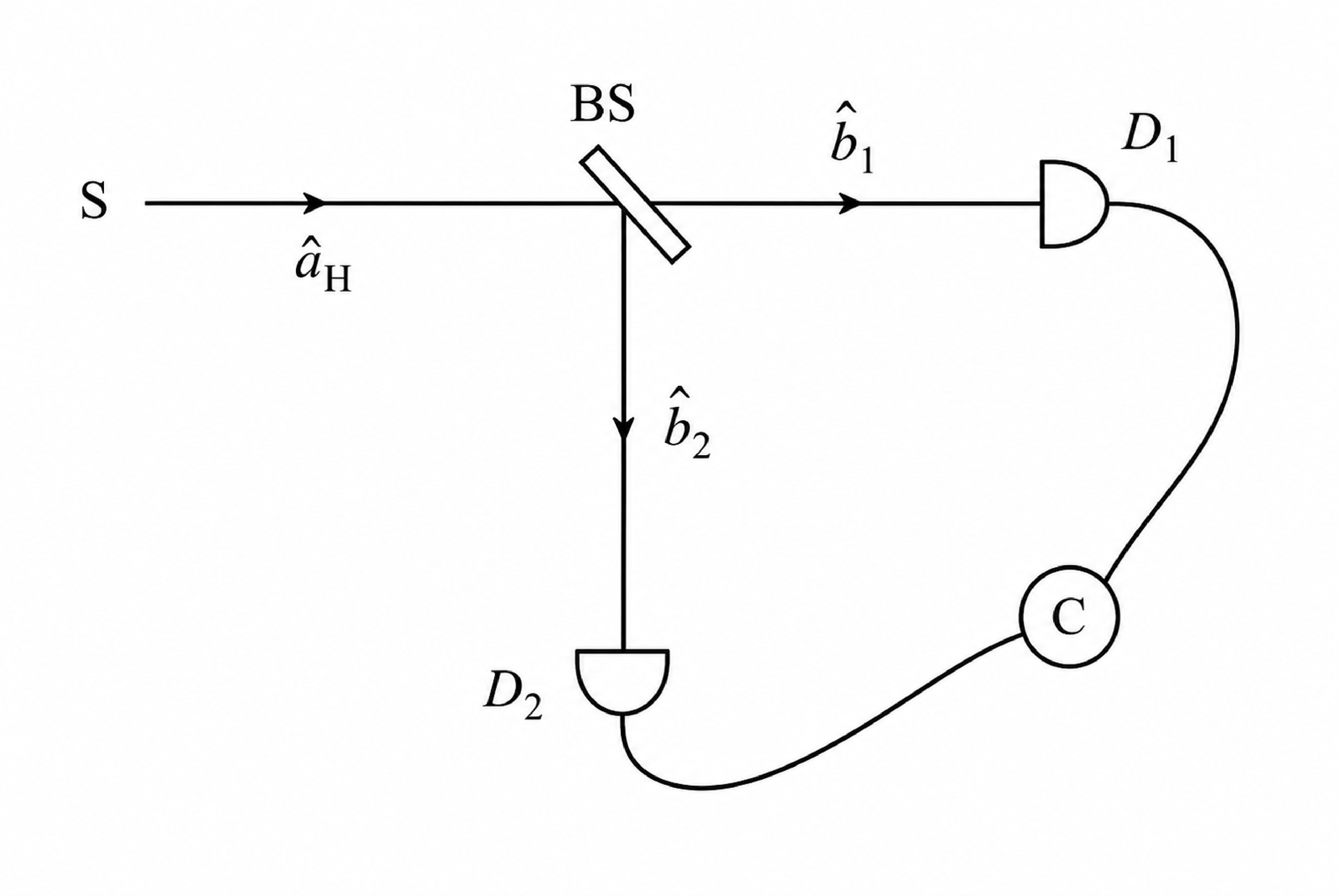}
    \caption{HBT configuration using a single-mode source ($S$) and two 
    	photo-detectors ($D_1,\, D_2$).}
    \label{fig:hbt_1mode}
\end{figure}

We now place a polarizing filter in front of a $2$-mode source such that 
it allows only one mode $\hat{a}_H$ to 
pass through as shown in Fig.~\ref{fig:hbt_1mode}. 

After the beam splitter action, the modes 
reaching the detectors are: 
\begin{equation}
    \label{eq:single_mode_beamsplitter}
    \hat{b}_1 = \frac{1}{\sqrt{2}} (\hat{a}_H - \hat{v}_H), \quad 
    \hat{b}_2 = \frac{1}{\sqrt{2}} (\hat{a}_H + \hat{v}_H),
\end{equation}
where $\hat{v}_H$ is a single polarization mode of the vacuum 
state that is input at the unused port of the beam splitter. For a single-mode,
\begin{align}
         g^{(2)}(\tau, \bm r) &= \frac{\langle : \hat{b}
         _1^{\dagger} \hat{b}_1 \hat{b}_2^{\dagger} \hat{b}_2 : 
         \rangle}{\langle \hat{b}_1^{\dagger} \hat{b}_1 \rangle 
         \; \langle \hat{b}_2^{\dagger} \hat{b}_2 \rangle} 
         \label{eqn:a_hbt_single_mode_space_time} \\ 
         &= \frac{\langle : \hat{a}_H^{\dagger} \hat{a}_H \hat{a}_H^{\dagger}
         	 \hat{a}_H : \rangle}{\langle \hat{a}_H^{\dagger} \hat{a}_H \rangle \;
         	  \langle \hat{a}_H^{\dagger} \hat{a}_H \rangle}
            \label{eqn:b_hbt_single_mode_space_time}
\end{align}

Note that in Eq.~\eqref{eqn:a_hbt_single_mode_space_time}, the 
expectation value is calculated for the state $\ket{\psi}_s 
\otimes \ket{0}$, where $\ket{\psi}_s$ is the state input from 
source $S$, while for 
Eq.~\eqref{eqn:b_hbt_single_mode_space_time}, it is calculated 
for $\ket{\psi}_s$ alone, similar to what was being considered 
for calculating the expectation values for all the previous 
cases.

Rewriting Eq.~\eqref{eqn:b_hbt_single_mode_space_time} to 
explicitly show its space and time dependence:

\begin{equation}
\label{eqn:hbt_single_mode_space_time}
         g^{(2)}(\tau, r) = \frac{\langle : \hat{a}
         _H^{\dagger}(t_1, \bm r_1) \; \hat{a}_H(t_1, \bm r_1) 
         \; \hat{a}_H^{\dagger}(t_2, \bm r_2) \; \hat{a}_H(t_2, 
         \bm r_2) : \rangle}{\langle \hat{a}_H^{\dagger}(t_1, 
         \bm r_1) \; \hat{a}_H(t_1, \bm r_1) \rangle \; \langle 
         \hat{a}_H^{\dagger}(t_2, \bm r_2) \; \hat{a}_H(t_2, \bm 
         r_2) \rangle}
\end{equation}
\noindent which for $r = 0$ (no path difference) and 
$\tau = 0$ (no time difference) becomes
\begin{equation}
\begin{split}
\label{eqn:hbt_single_mode_no_time_space}
        g^{(2)}(0,0) &= \frac{\langle : \hat{a}_H^{\dagger}(0,0) 
        \; \hat{a}_H(0,0) \; \hat{a}_H^{\dagger}(0,0) \; \hat{a}
        _H(0,0) : \rangle}{\langle \hat{a}_H^{\dagger}(0,0) \; 
        \hat{a}_H(0,0) \rangle \; \langle \hat{a}_H^{\dagger}
        (0,0) \; \hat{a}_H(0,0) \rangle} \\
        &= \frac{\langle \hat{a}^{\dagger} \hat{a}^{\dagger} 
        \hat{a} \hat{a} \rangle}{\langle \hat{a}^{\dagger} 
        \hat{a} \rangle^2} \quad \quad (\hat{a}_H(0,0) \equiv 
        \hat{a}) \\
        &= \frac{\langle \hat{n}(\hat{n}-1) \rangle}{\langle 
        \hat{n} \rangle ^2}
\end{split}
\end{equation}

The motivation behind calculating the intensity-intensity 
correlation function for $r = 0$ and $\tau = 0$ is that
$g^{(2)}(0,0)$ measures the coincidence counts at the two 
detectors. If light were a classical wave, then both detectors 
would click simultaneously, and if they do not then it indicates 
quantum nature of light, since a single quantum particle cannot 
be simultaneously detected by both detectors.

Thermal light has $g^{(2)}(0,0) = 2$ and is seen to exhibit
photon bunching, while  for coherent light $g^{(2)}(0,0) = 1$. 
In contrast, an ideal single-photon state satisfies 
$g^{(2)}(0,0) = 0$, indicating that it is impossible to 
simultaneously detect two photons for such sources, thus 
demonstrating perfect photon anti-bunching. In summary, 
$g^{(2)}(0,0)$ provides a direct signature of classical and 
non-classical radiation:
\begin{equation}
\begin{split}
    \label{eqn:g20}
    g^{(2)}(0,0) > 1 &: \text{Super-Poissonian photon statistics} \\
    & \quad \quad \implies \text{Photon bunching},\\
    g^{(2)}(0,0) = 1 &: \text{Poissonian photon statistics} \\
    & \quad \quad \implies \text{Coherent light}, \\
    g^{(2)}(0,0) < 1 &: \text{Sub-Poissonian photon statistics} \\
    & \quad \quad \implies \text{Photon anti-bunching}\\
    & \quad \quad \implies \text{Non-classical Behaviour}.
\end{split}
\end{equation}

While the HBT setup can help us measure anti-bunching of 
photons, a phenomenon that arises due to particle-like quantum 
behaviour of light; the question we ask here is, whether a 
similar setup can be used to unearth the non-locality 
associated with entangled states of light? To this end we 
replace the two photo-detectors in the HBT setup by two `on-
off' detectors. This allows us to measure the correlations of 
the type ``light-light'', ``nolight-light'' and ``nolight-
nolight''. The correlations between the binary possibilities at 
each detector can be described quantum mechanically by the 
measurement of an appropriate Hermitian operator with 
dichotomous eigenvalues. Further, we propose a toy-model of the 
HBT setup, which while keeping the essential structure 
simplifies the situation considerably and leading to the 
possibility of a clear and concise analysis. If we  demand that 
the correlations functions described above are arising out of a 
hidden variable model then they must satisfy an appropriate 
Bell-type inequality.

We analyze specific classes of quantum states of light and
show regions of parameters where they violate these Bell-type 
inequalities. Thus, in this work we show how HBT-kind of
setup can reveal quantum non-locality by the violations of
appropriate Bell-type inequalities.

\section{Bell-type inequality violations in HBT Toy-Model}
\label{sec:3_hbt_toy_model_detail}

Our proposed toy-model in Sec.~\ref{sec:2_2_hbt_toy_model}
demonstrates the HBT effect in a simple laboratory situation
and has been constructed with a view to explore
quantum effects. The toy-model is a valid counterpart of the
HBT setup as it has two sources and two detectors, and light
from each source can reach either of the detectors. For the
light-nolight correlations, we can setup a hidden variable
model and then show the violations of constraints emerging
from it to demonstrate quantum non-locality.

Consider first the typical HBT setup in Fig~\ref{fig:hbt_classical}. As explained earlier, light 
emitted from the two localized sources $S_1$ and $S_2$, is 
detected by two spatially separated detectors $D_1$ and 
$D_2$, located at a distance $L$ from the sources. Since 
each source contributes to the field detected at both 
detectors, there are four possible propagation paths 
connecting each source and detector in pairs, as depicted 
in the figure. With polarization included this corresponds
to an $8$-mode system.

We now describe the HBT toy-model proposed in Sec.~\ref{sec:2_2_hbt_toy_model} in greater detail. 
We construct our model by allowing the radiation from 
the two sources $S_1$ and $S_2$ to propagate along 
the $z$ and $z'$ directions, respectively. The two propagation directions 
are associated with orthogonal polarization bases $(x,y)$ and $(x',y')$, 
respectively. To reproduce the source-to-detector paths of the original HBT 
configuration, a suitable sequence of optical elements is placed along the 
two trajectories. The setup consists of four beam splitters, 
$B_k, \; k = 1, \ldots, 4$, two mirrors, $M_k$, and two linear polarizers, 
$P(\theta_m), \; m = 1, 2$, arranged as shown in Fig.~\ref{fig:hbt_toy_model}.

Each of the two sources generate $2$-mode entangled quantum resource 
states that are input to two balanced $(50-50)$ beam splitters $B_1$ and 
$B_2$. These states are separable from one another. The second input port 
of the beam splitters have $2$-mode vacuum states. The output from these 
beam splitters are then directed towards the second pair of beam splitters 
$B_3$ and $B_4$ using a specific arrangement of mirrors ($M_k$) to imitate 
the original HBT interferometer. The placement of the optical devices is 
such that each beam is always directed parallel to the others, thus confirming 
that we are confined to the paraxial regime.

Following this, two output modes from each beam splitter $B_3$ and $B_4$ which 
correspond to the source states, are retained for measurement, while the 
remaining two output modes corresponding to the vacuum states are discarded 
as redundant modes (indicated by red crosses in the toy-model schematic).
Before detection, each output beam passes through a linear polarizer 
$P(\theta_1)$ or $P(\theta_2)$, which selects the corresponding polarization 
component according to its orientation. The resulting fields are finally 
measured by the on-off detectors $D_1$ and $D_2$. Thus, the optical arrangement 
provides a $4$-mode realization of the $2$-source $2$-detector HBT geometry.

\subsection{Preparation of the Resource State}
\label{sec:3_3_hbt_toy_model_state}

For each of the two sources $S_1$ and $S_2$, we prepare a two-
mode squeezed vacuum (TMSV) state as the resource state. We 
begin with a general state of the $2$-mode radiation field 
with a centered Gaussian Wigner distribution, with $V = 
\frac{1}{2} G^{-1}$ as the covariance or noise matrix:

\begin{equation}
    \begin{split}
		\label{eqn:wigner1}
		&W_{1 (2)}(\xi_{1 (2)}) = \frac{1}{\pi^2} 
        (\text{Det} \, G_{1 (2)})^{1/2} \text{exp}(-\xi_{1 (2)}^T G_{1 (2)} \xi_{1 (2)}), \\
		&\xi_1^T = (q_1 \; p_1 \; q_2 \; p_2), \\
        &\xi_2^T = (q_3 \; p_3 \; q_4 \; p_4).
	\end{split}
\end{equation}

If the smallest eigenvalue of the covariance matrix $V$ for a 
given state is less than $\frac{1}{2}$, then the state is said 
to be squeezed and is therefore non-
classical~\cite{simon_1994}. In our work, we have applied the 
$2$-mode squeezing symplectic transformation $(Sq(\cdot))$ to 
obtain the matrix $G$ which renders the Wigner function in 
Eq.~\eqref{eqn:wigner1} to correspond to a TMSV state. Details 
on symplectic transformations are discussed in 
Appendix~\ref{sec:7_appendixA}. For sources $S_1$ and $S_2$ 
with squeezing parameters $u$ and $v$ respectively,

\begin{equation}
    \begin{split}
		\label{eqn:wigner2}
		  &G_1 = Sq(u)^{-1} \; G_0 \; Sq(u)^{-1}, \\
        &G_2 = Sq(v)^{-1} \; G_0 \; Sq(v)^{-1}, \\
        &G_0 = \kappa \; I_{4 \times 4}, \\
        &\kappa = \text{tanh} \frac{\hbar \omega}{2 k_B T}, 
        \quad 0 \le \kappa \le 1.
	\end{split}
\end{equation}

Here, $\kappa = 1$ indicates temperature $T = 0$ and $\kappa < 
1$ corresponds to some non-zero finite temperature. The 
squeezing symplectic transformation, which is an $Sp(4,
\mathcal{R})$ matrix, for $S_1$ is

\begin{equation}
    \begin{split}
        \label{eqn:squeezing_entanglement}
        &Sq(u) = \begin{pmatrix}
		\text{cosh}(u)\; I_{2 \times 2} & 
        \text{sinh}(u)\; Z_{2 \times 2}\\
		\text{sinh}(u)\; Z_{2 \times 2} & 
        \text{cosh}(u)\; I_{2 \times 2} \\
		\end{pmatrix}, \\
        &Z_{2 \times 2} = \begin{pmatrix}
		1 & 0\\
		0 & -1
        \end{pmatrix}.
    \end{split}
\end{equation}

Similarly, we can construct the corresponding Wigner 
distribution for $S_2$ with squeezing parameter $v$. The 
prepared $2$-mode TMSV states are sent in as inputs from $S_1$ 
and $S_2$ to the HBT toy-model. Initially, each of them is an 
entangled state but is separable from the other. They get 
transformed after they emerge out of the HBT optical system, 
such that now all four modes become entangled which leads to 
violation of the Bell-type inequality, as we shall see in 
Sec.~\ref{sec:5_results}.

\subsection{Bell-type inequality via local hidden variables}
\label{sec:4_bell_inequality}

In this section, we derive the Bell-type inequality that we 
employ in our work to show non-locality in our HBT toy-model. 
Our aim is to measure the coincidence count rates of photons 
incident on the two detectors. $D_1$ and $D_2$ are on-off 
detectors which perform coarse-grained measurements to 
distinguish `light' from `no light'.

We define the following four dichotomous hermitian operators 
having eigenvalues $0$ and $1$:

\begin{equation}
    \begin{split}
    \label{eqn:bell_operators}
    \hat{H}_1 &= (\mathcal{I}_{2 \times 2} - 
    \ket{00}\bra{00})_{S_1}, \\
    \hat{H}_2 &= (\mathcal{I}_{2 \times 2} - 
    \ket{00}\bra{00})_{S_2}, \\
    \hat{H}_1 (\theta_1) &= (\mathcal{I}_{\theta_1} - 
    \ket{0}_{\theta_1 \theta_1} \!\! \bra{0}) \, \mathcal{I}_{\theta_1 + \frac{\pi}{2}}, \\
    \hat{H}_2 (\theta_2) &= (\mathcal{I}_{\theta_2} - 
    \ket{0}_{\theta_2 \theta_2} \!\! \bra{0}) \, \mathcal{I}_{\theta_2 + \frac{\pi}{2}}.
    \end{split}  
\end{equation}

Here, $\hat{H}_1$ operates on modes $\hat{a}_1$ and $\hat{a}
_2$ when the polarizer $P(\theta_1)$ is removed from the toy-
model in Fig~\ref{fig:hbt_toy_model}, while $\hat{H}_1 
(\theta_1)$ is the corresponding operator when the polarizer 
is placed at an angle $\theta_1$. The unit operator for modes 
$\hat{a}_1$ and $\hat{a}_2$ in the absence of the polarizer is 
$\mathcal{I}_{2 \times 2}$, while that for the single mode 
propagating out of $P(\theta_1)$ at a polarization angle of 
$\theta_1$ is $\mathcal{I}_{\theta_1}$, and $ \mathcal{I}
_{\theta_1 + \frac{\pi}{2}}$ is for the orthogonal 
polarization. Similarly, we have defined operators $\hat{H}_2$ 
and $\hat{H}_2 (\theta_2)$ which act on both modes $\hat{a}_3$ 
and $\hat{a}_4$ in the same way as explained above. Note that 
the form of the operators $\hat{H}_1 (\theta_1)$ and $\hat{H}
_2 (\theta_2)$ is such that they already have the action of 
the polarizers ingrained in them. 

Expectation values of these measurement operators gives the 
corresponding probabilities. For example, $\langle \hat{H}_{1 
(2)} \rangle$ gives the probability of detecting at least one 
photon at $D_{1 (2)}$ with polarizer $P(\theta_{1 (2)})$ 
absent. Similarly, $\langle \hat{H}_{1 (2)} (\theta_{1 (2)}) 
\rangle$ gives the probability of detecting at least one 
photon at $D_{1 (2)}$ with polarizer $P(\theta_{1 (2)})$ fixed 
at an angle $\theta_{1 (2)}$.

Let $h_{1 (2)} (\lambda), \; h_{1 (2)} (\theta_{1 (2)}, 
\lambda)$ be the actual values of the dynamical variables 
$H_{1 (2)}, H_{1 (2)} (\theta_{1 (2)})$ for a given value of 
the hidden variable $\lambda$ respectively. The average values 
of the products of the operators in 
Eq.~(\ref{eqn:bell_operators}), corresponding to the various 
coincidence count rates, are then given in local hidden 
variable theory by

\begin{equation}
    \begin{split}
        \label{eqn:hvt_integral}
        E(\theta_1, \theta_2) &= \langle H_1 (\theta_1) \; H_2 (\theta_2) \rangle_{hv} \\
        &= \int h_{12} (\theta_1, \theta_2, \lambda) \; \mathcal{P}(\lambda) \; d\lambda \\
        &= \int h_1 (\theta_1, \lambda) \; h_2 (\theta_2, \lambda) \; \mathcal{P}(\lambda) \; d\lambda
    \end{split}
\end{equation}

\noindent where $\mathcal{P}$ is the probability distribution 
for $\lambda$. The dependence of $\mathcal{P}$ on the chosen 
quantum state is neglected for simplicity. We write
$h_{12} (\theta_1, \theta_2, \lambda)$ as product of two 
factors $h_1 (\theta_1, \lambda)$ and $h_2 (\theta_2, 
\lambda)$, each term being dependent on only one polarizer 
angle. This follows from the locality condition of the HBT 
model explained in Sec.~\ref{sec:3_hbt_toy_model_detail}. 
Similarly from Eq.~\eqref{eqn:hvt_integral}, we can write 
expressions for the other correlation functions: 
$E(\; ,\; ) = \langle H_1 H_2 \rangle, \; E(\;, \theta_2) =
\langle H_1 H_2 (\theta_2) \rangle$ and $E(\theta_1, \;) = 
\langle H_1 (\theta_1) H_2 \rangle$.
  
Using the Clauser-Horne lemma and the ``no enhancement 
assumption" (i.e, $h_i(\theta_i, \lambda) \le h_i(\lambda)$), 
we can derive the following inequality:

\begin{equation}
\label{eqn:bell_inequality}
    \begin{split}
        -E(\;, \;) \le \, &E(\theta_1, \theta_2) - 
        E(\theta_1, \theta_2') + E(\theta_1', \theta_2) \\
        &+ E(\theta_1', \theta_2') - E(\theta_1', \;) - 
        E(\;, \theta_2) \le 0.
    \end{split}
\end{equation}

This is the state-independent generalized Bell-type inequality
for arbitrary multi-photon states. The no-enhancement 
assumption that was used here states that introducing a 
polarizer cannot increase 
the detection probability compared with the case when the polarizer is absent. Detailed derivation of 
this inequality can be found in~\cite{arvind_1999}. Violation 
of this inequality by a given state using quantum theory, 
indicates that the state has non-local quantum correlations 
that cannot be explained by a realist local hidden variable 
model. Now that we have a hidden variable description, 
we shall work with Eq.~\eqref{eqn:bell_inequality} to show 
non-locality in our HBT toy-model in the next subsection.

\subsection{Non-locality in HBT Toy-Model}
\label{sec:5_results}

For the particular class of centered Gaussian Wigner states, 
defined in Sec.~\ref{sec:3_3_hbt_toy_model_state}, which we 
feed as input resource states to our HBT toy-model, we check 
for violations of the Bell-type inequality 
(Eq.~\eqref{eqn:bell_inequality}) of the transformed states 
after they emerge out from the system, by calculating

\begin{equation}
\label{eqn:bell_function_qm}
    \begin{split}
        f(\theta_1, \theta_2, \theta_1', \theta_2') &=
        E(\theta_1, \theta_2)_{qm} - E(\theta_1, \theta_2')_{qm} \\ 
        & \quad + E(\theta_1', \theta_2)_{qm} + E(\theta_1', \theta_2')_{qm} \\
        & \quad \quad - E(\theta_1', \;)_{qm} 
        - E(\;, \theta_2)_{qm}.
    \end{split}
\end{equation}

\begin{figure}
    \centering
    \subfloat[\centering Plot for Bell observable 
    $f(\theta_1, \theta_2, \theta_1', \theta_2')$ against squeezing 
    parameter $u$, considering the three possible relations 
    between $v$ and $u$, i.e., $v = 0, v = u$ and $v = -u$, for polarizer angles $\theta_1 = \frac{\pi}{8}, \theta_2 = \frac{\pi}{4}, 
    \theta_1' = 3\frac{\pi}{8}, \theta_2' = 0$ and $\kappa = 1$.]{{\includegraphics[width=\linewidth]{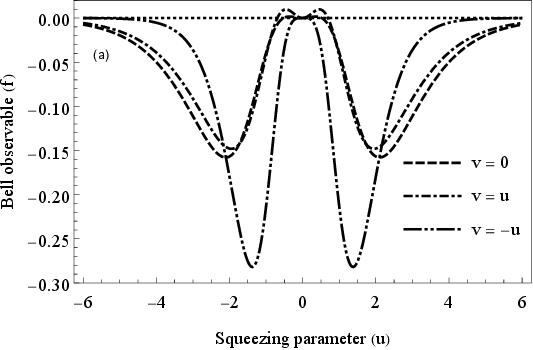} }}
    \\ 
    \subfloat[\centering Enlarged portion of plot (a) to clearly
    show the range of $u$ over which the inequality is violated.]
    {{\includegraphics[width=\linewidth]{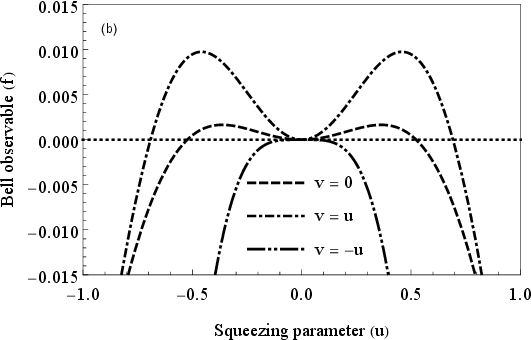} }}
    \caption{Bell observable $f(\theta_1, \theta_2, \theta_1', \theta_2')$ as a function of the squeezing parameter $u$ for two $2$-mode squeezed-vacuum (TMSV) resource states emerging from the HBT toy-model, illustrating the violation of the Bell-type inequality. Here, $u$ and $v$ are the squeezing parameters of the two sources.}
    \label{fig:hbt_plots}
\end{figure}

We search over all values of polarizer angles 
$\theta_1, \theta_2, \theta_1', \theta_2'$ to locate possible 
violations of the inequality~\eqref{eqn:bell_inequality}. In 
Fig~\ref{fig:hbt_plots}, we show violations for $\kappa = 1$ 
and special polarizer angles $\frac{\pi}{8}, \frac{\pi}{4}, 
3\frac{\pi}{8}, 0$, for three combinations of the squeezing 
parameters $u$ and $v$.

We can see from Fig~\ref{fig:hbt_plots} that the Bell-type 
inequality in Eq.~\eqref{eqn:bell_inequality} is clearly 
violated for a particular range of squeezing parameter $u$. The 
plots include all three cases through which $v$ and $u$ can be 
related, i.e., $v = 0, v = u$ and $v = -u$. At $u = v = 0$, we 
have $f = 0$, indicating that the inequality is satisfied when 
there is no squeezing, i.e., when both input states are 
classical states, there is no violation. Violations are 
apparent for squeezing values close to zero and die down as 
squeezing is increased. The plots show maximum violation for 
the case when $v = u$, i.e., when all modes are squeezed (modes 
$1, 3$ by $u$ and modes $2, 4$ by $-u$). Surprisingly, we also 
see that for $v = -u$ there is no violation, even though this 
too corresponds to the case when all modes are squeezed (modes 
$1, 4$ by $u$ and modes $2, 3$ by $-u$). When $v = 0$, i.e., 
when modes corresponding to $S_1$ are squeezed by equal and 
opposite amounts (entangled state), while those corresponding 
to $S_2$ do not undergo any squeezing (classical state), we see 
a small violation of the Bell-type inequality.

\begin{figure}
    \centering
    \includegraphics[width=\linewidth]{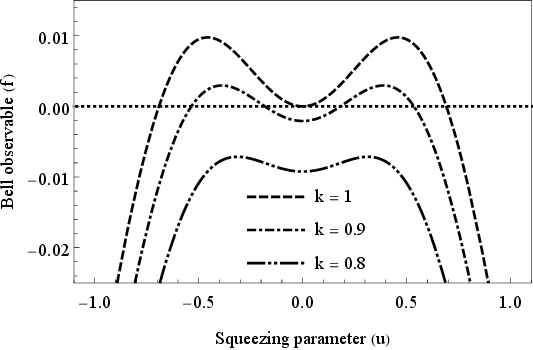}
    \caption[width=0.5\textwidth]{Plot for Bell observable 
    $f(\theta_1, \theta_2, \theta_1', \theta_2')$ and squeezing 
    parameter $u$ with $v = u$ for squeezed thermal states 
    ($\kappa < 1$.)}
    \label{fig:hbt_k_plot_zoom}
\end{figure}

In Fig~\ref{fig:hbt_k_plot_zoom}, we plot the Bell observable 
$f(\theta_1, \theta_2, \theta_1', \theta_2')$ against squeezing 
parameter $u$, for the case $v = u$, for squeezed thermal 
states, i.e., $\kappa < 1$. When the initial state is squeezed 
thermal, the amount of violation gradually decreases with 
decrease in $\kappa$, as compared to when it is a squeezed 
vacuum state ($\kappa = 1$). Hence, we see that even at finite 
temperatures, there are traces of non-local correlations, which 
gradually disappear with increase in temperature or decrease in 
$\kappa$.

\section{Conclusion}
\label{sec:6_conclusion}

The HBT effect, credited to be a breakthrough experiment, 
initiated the notion of quantum nature of light that eventually 
led to the development of the field of `quantum optics' in a 
way that is known to us today. In this work, we have proposed a 
toy-model for the HBT setup, but instead of its usual $8$-
polarization mode description, we have simplified our model so 
as to work with only $4$-modes, leading to a simpler 
mathematical analysis of our optical system. We have shown that 
the toy-model is a logical representation of the typical HBT 
setup in the sense that it satisfies the locality condition 
required for a classical hidden variable theoretic description.

Further, we prepare two TMSV resource states which are sent as 
inputs to our toy-model. We show that the output states violate 
the Bell-type inequality at small squeezing values. We have 
studied Bell violations for squeezed vacuum states 
$(\kappa = 1)$ and squeezed thermal states $(\kappa < 1)$. The 
toy-model entangles the two TMSV states that were originally 
separable from one another, which was the reason behind the 
violations. We also conclude that if we have one non-classical 
input state, the other input state being a classical one, the 
HBT setup still entangles the outputs leading to violations. 
Only if both states are quantum optically classical, we do not 
observe violations.

In this work, the detectors are assumed to have the capacity to
distinguish `light' from `no-light'. A possible future 
direction could be to refine our measurements further for a 
detailed analysis of non-locality.

\vspace*{0.5cm}

\section*{Acknowledgement}
\label{sec:acknowledgement}

S.C. acknowledges the Prime Minister’s Research Fellowship 
(PMRF) scheme, GoI, for financial support.

\vspace*{0.3cm}

\appendix

\section{Continuous-Variable (CV) Systems and Symplectic 
transformations}
\label{sec:7_appendixA}

There are two possible ways in which an $n$-mode CV system can 
be described. One is using $n$ pairs of conjugate quadrature 
field operators, $\hat{q}_j, \hat{p}_j$, $(j = 1,2,\dots,n)$, 
and the other is using $n$ pairs of annihilation and creation 
operators, $\hat{a}_j, \hat{a}_j^{\dagger}, (j = 1, 2, \dots, n)$.

The quadrature operators can be arranged into a column vector 
$\hat{\xi} = (\hat{q}_1 \; \hat{p}_1 \; \hat{q}_2 \; \hat{p}_2 
\; \dots \; \hat{q}_n \; \hat{p}_n )^T$, which satisfy the 
canonical commutation relation:

\begin{equation}
	\label{eqn:6_commu_relations}
	[\hat{\xi}_{i}, \hat{\xi}_{j}] = i \; \Omega_{ij}, 
    \quad 
	\Omega = \bigoplus_{i=1}^n \; \begin{pmatrix}
		0 & 1 \\
		-1 & 0
	\end{pmatrix}.
\end{equation}

They are related to the annihilation and creation operators as 
$\hat{q} = (\hat{a} + \hat{a}^{\dagger})/\sqrt{2}$ and $\hat{p} 
= i (\hat{a}^{\dagger} - \hat{a})/\sqrt{2}$.

For an $n-$mode system, the Wigner function corresponding to a 
density operator $\hat{\rho}$ is defined as

\begin{equation}
	\label{eqn:6_wigner_def}
	W(\xi) = \frac{1}{(2 \pi)^{n}} \int d^n x' \; \Big\langle
    x - \frac{1}{2} x' \; \Big| \; \hat{\rho} \; \Big| \; x 
    + \frac{1}{2} x' \Big\rangle \; \text{exp}(i x' \cdot y), 
\end{equation}

\vspace{0.3cm}

\noindent where $x = (q_1 \; q_2 \; \dots \; q_n)^T$ and 
$y = (p_1 \; p_2 \; \dots \; p_n)^T$.

For an $n-$mode system, the first-order moment is the 
displacement vector and is given by 
$\textbf{d} = \langle \hat{\xi} \rangle = Tr[\hat{\rho} 
\hat{\xi}]$. The real symmetric $2n \times 2n$ covariance 
matrix forms the second-order moment and is given by 
$V = (V_{ij}) = \frac{1}{2} \langle \{ \Delta \hat{\xi}_i, 
\Delta \hat{\xi}_j \} \rangle$, where 
$\Delta \hat{\xi}_i = \hat{\xi}_i - \langle \hat{\xi}_i 
\rangle$ and $\{ , \}$ is the anti-commutator.

For a Gaussian state with displacement \textbf{d} and 
covariance matrix $V$, the Wigner function in 
Eq.~\eqref{eqn:6_wigner_def} reduces to
\begin{equation}
	\label{eqn:6_wigner_gaussian}
	W(\xi) = \frac{\text{exp} [-(1/2) \; (\textbf{$\xi$} 
    - \textbf{d})^T \; V^{-1} \; (\textbf{$\xi$} - \textbf{d})]}
    {(2 \pi)^n \sqrt{\text{det} V\;}}.
\end{equation}

\bibliography{references}

\end{document}